# Sustaining Research Software: an SC18 Panel

Daniel S. Katz[1], Patrick Aerts[2], Neil P. Chue Hong[3], Anshu Dubey[4], Sandra Gesing[5], Henry J. Neeman[6], David E. Pearah[7]

[1] NCSA & CS & ECE & iSchool, University of Illinois Urbana-Champaign, d.katz@ieee.org
[2] Netherlands eScience Center and Data Archiving and Networked Services, p.aerts@esciencecenter.nl
[3] Software Sustainability Institute, EPCC, University of Edinburgh, N.ChueHong@software.ac.uk
[4] Argonne National Laboratory and University of Chicago, adubey@anl.gov
[5] Center for Research Computing, University of Notre Dame, sandra.gesing@nd.edu
[6] University of Oklahoma, hneeman@ou.edu
[7] The HDF Group, david.pearah@hdfgroup.org

## Introduction

Many science advances have been possible thanks to the use of research software, which has become essential to advancing virtually every Science, Technology, Engineering and Mathematics (STEM) discipline and many non-STEM disciplines including social sciences and humanities. And while much of it is made available under open source licenses, work is needed to develop, support, and sustain it, as underlying systems and software as well as user needs evolve.

In addition, the changing landscape of high performance computing (HPC) platforms, where performance and scaling advances are ever more reliant on software and algorithm improvements as we hit hardware scaling barriers is causing renewed tension between sustainability of software and its performance. We must do more to highlight the trade-off between performance and sustainability, and to emphasize the need for sustainability given the fact that complex software stacks don't survive without frequent maintenance; made more difficult as a generation of developers of established and heavily-used research software retire. Several HPC forums are doing this, and it has become an active area of funding as well.

In response, the authors organized and ran a panel at the SC18 conference. The objectives of the panel were to highlight the importance of sustainability, to illuminate the tension between pure performance and sustainability, and to steer SC community discussion toward understanding and addressing this issue and this tension. This panel was intended to have greater audience participation than a typical SC panel. In addition to presentations by the panelists, and questions from the audience to the panel, we used interactive polling to gather audience inputs and to guide discussion between the panelists as well as with the audience.

The outcome of the discussions, as presented in this paper, can inform choices of advance compute and data infrastructures to positively impact future research software and future research.

## Format

The panel started with a 5 minute introduction from the moderator to define the problem of software sustainability in a research context. Panelists then provided their insights and/or relevant experiences in 5 minute talks followed by a quick, moderated, full-group discussion of the talks. This was followed by online polling through a platform where the results of input were displayed instantaneously. This strategy was very successful in engaging the audience and the outcome of the exercise is summarized later in the report.

## Panelists

Panelists included globally acknowledged leaders in the field of software sustainability. One of the panelists leads an institutional HPC center working with consumers of research software, one leads a center that both builds and uses research software, and the others are key members of relevant projects, organizations, and efforts at the vanguard of sustainable software. For example, sustainability is a key challenge in the US Department of Energy's (DOE's) [Exascale Computing Project (ECP)](), with the [Interoperable Design of Extreme-scale Application Software (IDEAS)]() project and the [Better Scientific Software (BSSw)]() effort as represented by panelist Dubey seen as a path for progress in this area; in the Cyberinfrastructure for Sustained Scientific Innovation (CSSI) program in the US National Science Foundation (NSF), represented by moderator Katz and panelist Gesing who are co-PIs in the US Research Software Sustainability Institute (URSSI) Conceptualization project; in the UK, represented by panelist Chue Hong as the director of the national [Software Sustainability Institute (SSI)](); in the Netherlands, represented by panelist Aerts, combining his work at the [Netherlands eScience Center](), where sustainability is a key goal for the software produced by the center and at DANS, [Data Archiving and Networked Services](), where the emphasis is also on sustaining software from the cultural heritage point of view; and in companies such as the [HDF Group](), represented by panelist Pearah (CEO, [HDF Group]()), where the company's future depends on the sustainability of the software products they build. Anyone who either develops or uses software should be concerned about its sustainability.

The panelists have taken their previous experience, as software developers, users, integrators of complex software applications and systems, and service providers, to generalize what they have learned about sustainability and working with larger communities to collectively understand the state of the art and make improvements, such as in [SSI](), [URSSI](), [WSSSPE](), [IDEAS](), [BSSw](). In addition to these overall projects, they have also worked together in smaller groups.

**Sandra Gesing** introduced aspects of software sustainability via a 2017 [survey]() of members of the National Postdoctoral Association. Results include that 95% of US-based researchers and 90% of UK-based researchers (from a prior, similar [survey]()) answered that they use software for their research. 63% (US) or 70% (UK) cannot perform their research without software. Given these numbers and that the people surveyed also often create their own software, it is alarming that over 50% lack formal or informal training in software engineering. While the UK already achieved a career path for [Research Software Engineers (RSEs)]() via the initiatives organized by the UK [Software Sustainability Institute (SSI)]() and the [RSE Association](), the US still lacks well-defined career paths.

Software sustainability has gained increased attention in academia, and this deficit in career paths in the US academic landscape has been recognized as a major challenge. Quite a few projects and initiatives have been funded or formed in the US with diverse foci and approaches that include improving career paths for RSEs, developers, facilitators, research programmers as one of their goals - to name a few: URSSI, SSI, [Science Gateways Community Institute (SGCI)](), national RSE Associations forming in [Germany](), the [Netherlands](), and the [US]() as well as [ACI-REF (Advanced Cyberinfrastructure – Research and Education Facilitators)]() and its successor [CaRCC (Campus Research Computing Consortium)](). All such projects recognize improving career path as a long-term goal, since it requires a cultural change in academia and a system incentivizing software development and facilitation to support research.

The diverse list of projects shows the interest in changing the culture in academia to support career paths, but the interaction between these initiatives is still sparse. Many PIs, senior personnel, and organizers of these projects are also involved in at least one other project, but there is no systematic way for such organizations to collaborate. The implementation of URSSI would aim at developing such a systematic way, provide successful use cases for career paths at single universities, e.g. at [Notre Dame]() and [Princeton](), analyze how this worked in the UK to understand US similarities and differences, and would aim at improving access to tools and standards that help universities hire diverse talent.

**Patrick Aerts** started out by dividing the issue in two major parts, with something in the middle: the past (legacy), the future (just easy to maintain codes), and in the middle, the present where we have to work on insights on how to create software in a sustainable manner and provide education about this. He also presented three take-home massages to get things done:
1. Treat Software Sustainability and Data Stewardship on an equal footing, at least policy-wise. Seriously consider linking up with RDA, under a separate chapter.
2. Consider Software (and data) as value objects. Then it starts making sense to spend some to keep the value or increase it.
3. Make the stakeholders' positions explicit, define their role, and involve all of them. Distinguish the interests of funders, scientists, and executive organizations.

His contribution stressed the importance of broadening the discussion on legacy to the whole cultural sector: (national) libraries, (national) archives, digital born or digital supported art, games, vision and sound collections, and science of course.

While it is understandable that in the HPC-domain, most of the interest is in best guidelines for newly-to-be-designed software, the topic as a whole involves all the mentioned domains, because many of their data did became unreachable, due to software that has stopped working on new platforms. His estimate is that this involves mainly data from the 1990s and early 2000s. A few examples.
- The games department: Communities are formed many keeping or reviving games.
- In art: older digital born artifacts suffer from obsoleted platforms and physically stopped working. But one wants the arty experience to remain operational.
- The Dutch National Library keeps 15.000 CD-ROMs. Hardware to read these is hardly available, so images were extracted from the CD-ROMs (but thou shall not copy), mostly running under old operating systems (W95 or older), for which no licenses can be obtained.

- Sometimes obsolete hardware is the only means to recover old data by using the old software and start migrating from there (MS Word 4 or MacWrite, 400k or 800k Apple-diskettes, just no name a few) This requires lists with coordinates of still working old machinery.
- Compared to the challenges in the arts and gaming domains, reviving older scientific software seems relatively easy.

His advice for the software legacy domain is *to get all forces joined* into one or more experts networks, at an European, cross-Atlantic, or global level, and to start exchanging practical experiences on how to solve specific problems.

It remains to agree on smart guidelines for developing new software. The Netherlands are working on a web portal with the working title Software Deposit Route, with guidelines for what to do with newly written academic software. The purpose of this effort is to make sure that no matter which academic organization you ask the question of what to do with your software, you will get the same advice. It will also contain directions for writing software properly if you have not yet begun the coding. The eScience Center in the Netherlands also hosts a Research Software Directory, with indexes to software with a proven level of quality. Also various European ESFRI-type of EC-funded projects are designing their own software quality directives, which need to be harmonized. And finally we are happy to send people to the knowledgebase of the Software Sustainability Institute in the UK, which is basically ahead of many if not all, to document best practices and practical advise on the matter.

**Anshu Dubey** presented her insights from working with the FLASH (http://flash.uchicago.edu/site/flashcode/) code, which is composable multiphysics software designed for simulating phenomena modeled with partial differential equations. The code has been in existence for nearly two decades, and has grown from being a code for astrophysics to serving six or more diverse research communities. All through its existence FLASH has balanced between performance on one side, and flexibility, extensibility,and sustainability on the other. Many of challenges faced by FLASH are common to all research software. These challenges are:
- Because the real world is messy, obtaining modularity and encapsulation in software modeling can be particularly difficult.
- As scientific understanding grows, so does the complexity of the code.
- With increasing platform heterogeneity, and increasing heterogeneity in solvers, there are now two orthogonal axes of complexity that codes must contend with.
- All aspects of software development can be under research, including model, algorithms, numerical techniques etc.
- Because teams developing such codes tend to have people with diverse training and backgrounds, constructive interdisciplinary interactions are necessary, but can be challenging.
- Incentive structures are not the same for all team members, which can lead to tension for prioritization.
- Almost all projects compete for resources both internally and externally.

**David E. Pearah** presented his perspective on why "scientific" software faces unique sustainability challenges as compared to "regular" software: the incentives are different, thus motivations and efforts would be as well. He brought up three main issues:

- Scientific software seeks minimum **run**-time time while regular software seeks minimum **code authoring** time
- Scientific software's main constraint is HPC cycles (hardware) while regular software's main constraint is programmers (both availability and cost)
- The timeframe for the life of scientific software is often just to support a single person to publish a paper, whereas regular software's goal is to support lots of strangers (forever)

Like many open source companies—particularly in the scientific domain—consulting is the only financial engine to support the programmers, but nearly all new funding is tied to new features and functionality, not software quality or maintenance. Also similar to many, though not all, open sources companies, another avenue to seek sustainable investment is to create an "Enterprise" or "Pro" version of the otherwise freely available software. This can sometimes be seen as controversial since it goes against the spirit of open and free exchange. Pearah proposed that a sustainable economy for open source needs to be created, simply starting with showing appreciation and consideration for those projects that we all rely on every day.

**Neil P. Chue Hong** presented his perspective on why we should be aiming to write less software to improve the sustainability of research software, based on experiences from the Software Sustainability Institute and studies done by the SSI and others. He observed that a lot of people write software, there is a wide variety of software that is used in practice, and that much software that is referenced in papers is never updated after publication. To address these issues, he proposed that we should, as a community, be "incentivising researchers to write the smallest possible amount of new code" through measures such as encouraging them to build on and extend existing software / platforms (e.g., through grant guidelines, making it easier to discover software that meets needs); to treat these platforms as infrastructure, and fund maintenance as such (e.g., through "taxation"); and to use this extra funding to enable research software engineers to focus care on established software, as well as consulting on new software. Overall, the challenge is to balance the desire to explore novel functionality with the requirement to consolidate and improve software so that it can be scalably supported.

**Henry J. Neeman** presented thoughts on containers and their contribution to software sustainability. In particular, he discussed the near term positive value of containers as a mechanism to ensure software portability and therefore scientific reproducibility, but also the likely medium to long term consequences associated with, ironically, the fact that making the software more straightforward to port to new platforms can be expected to lead to a reduction in software performance relative to (a) new platform capability and (b) growing run sizes.

Especially as hardware performance of later generations improves, some of these performance improvements are unavailable to older containerized applications, because those containers aren't ported to new compiler versions that can exploit the new CPU capabilities (e.g., AVX/2/512 vs SSE1/2/3/4 vector instructions), and older compiler versions are much less likely to be updated to be able to exploit these new CPU capabilities. For codes that are entirely memory bound and that therefore gain no value from new CPU capabilities, this may be acceptable, but for other codes, especially those that are at least partially CPU-bound (arguably most STEM research codes), it will become increasingly difficult for researchers to keep up with exponentially growing data and run sizes. At the same time, because containerization makes porting (at all) to new platforms much more straightforward, it disincentivizes porting directly onto these new platforms (that is, creating a new container with a new operating system, with software built by a new compiler version, etc). And, the longer a software developer puts off this porting, the more difficult the porting becomes, further disincentivizing the increasingly difficult porting task.

In December 2018, Neeman published an invited blog post at the URSSI website that provided greater detail on these ideas: http://urssi.us/blog/2018/12/21/why-research-software-sustainability-wont-be-fixed-by-containers/

## Audience Input

After the panel presentations, we used an interactive polling platform ([mentimeter.com](mentimeter.com)) to get feedback from the audience on a number of key questions, listed below.

We first asked if the audience members were here to learn about sustainability in general or because they had a specific software project that has sustainability challenge. 41 said to learn about sustainability in general, and 22 said because of a specific project.

Next, we asked what sustainability meant to the attendees. Results were:

- The software is ready to do the next new requirement at all times
- The software should be maintained and improve over time
- Making it possible to use and improve software over a long period of time
- Maintaining updating improving using
- Not proof of concept
- Better programming standards
- Long term support and maintenance
- Community
- Software engineering
- Works with changing architecture
- Have software that runs now and in the future
- Used more than 5 years later
- Software that is useful for the long haul - 20-30 years?
- Support to make it work somehow long term
- Ability to maintain software at a great level
- Long term support for software users
- Being able to keep software usable for existing users while being able to add new users
- The ability for future users to understand and use a code
- Someone, sometime in the future will be able to reuse and adapt said software
- Keeping software relevant with minimal manpower effort
- Shareable, portable, understandable by future developers
- Keeping older software functioning on newer systems
- Users can rely on it being maintained, or can contribute via an active community
- Being able to use the software on multiple architectures and be able to use the code for the foreseeable future
- The nice model we implemented 5 years ago will still run in 5 years time
- Making software that will be usable and useful for the next several years

- Ensuring the long term viability and use
- Software that is future proof
- Applying principles for good design in software engineering to scientific codes
- Software updates to keep up with hardware / compiler / library changes
- Continued availability of functioning software and access to support
- Code not written incurs no maintenance
- Robust to changes in technology
- Software that has at least the same features as N years ago
- Reliable, stable code that others can use more than once
- A way to keep a codebase relevant and useful over time
- Keeping useful software useful
- The ability to improve software in both an economical and timely manner
- Longevity, reliability
- Software is maintained and incremental development is supported to address ongoing needs of the users and adaptation to new architectures
- Software is usable and easy to maintain as long as needed
- a) Being able to re-run my experiment, and b) being able to run a new experiment with the same software
- Extensible and easy to update
- Ease of use for new adopters. Backwards compatibility. Low transactional cost for development
- As little cost as possible to port to new architectures
- It doesn't give me a heart attack when I hear "porting"
- Maintaining value, improvability/flexibility, accessibility of code
- Reusable, lasts for at least 10 years, bug free development, documented, easy to use
- Software/hardware independence/resilience/use/maintenance over time
- Software that continues to meet the needs and expectations of users long term such that they do not abandon it
- Create, use, maintain research software. The challenge is software fixes, maintenance, portability, and enhancements. People is key
- The cost of changing software shouldn't grow nonlinearly over time

The common points in these suggestions are longevity and adaptability to changes (in user needs, platforms, etc.)

We then asked attendees for an example of software that they consider sustainable, and the top reason they think it is sustainable, and were told:

- BLAS--provides framework/building blocks for architecture independent optimization
- Jupyter, it's well funded
- Root, a large domain specific community uses it and increasingly other communities. Maybe CERN has something to do with its sustainability.
- Emacs
- Python, because it has thriving communities of both users and maintainers
- LLVM. Corporate support $$$
- Linux - because there is a community contributing to it
- Nothing indefinitely

- LLVM due to large, active community
- LAPACK - it still forms the core of so many applications and libraries
- Microservices - small software delivering specific functions
- Intel MKL, emacs, Linux, iOS
- Git
- Python
- Lapack - well established interfaces and strong testing
- Fortran It's still around
- Linux - broad usage, diverse use cases, very large community, many businesses
- nixpkgs
- Quantum espresso - community
- Linux. It has a large user community, including developers and organizations contributing resources (money and time) to continue development and support
- c/c++: fundamental usage
- MPI libraries. FFTW. Compilers - gcc etc.
- Gcc
- Python: large enough community to get the snowball rolling pretty well on its own.
- MPI stacks because of many users and outside interest
- Vtk (visualization toolkit) is an open source library developed by the community that has been developed for over 20 years and has had long term funding to sustain it
- C/C++. Because of community engagement, broad usage and support
- OpenMPI - modular, well-designed, strong community
- Many of the UK's Collaborative Computing Projects (CCPs): Long term stable funding and stable development teams (not just PhD students). Good software engineering.

One common element among these responses are projects that are generally very widely used, across disciplines, or in other words, those who had large communities who could sustain them. Another common element is long-term funding, either from grants or from industry. A final element is software engineering or code organization that makes sustainability more likely.

Next, we asked the attendees if their community considers sustainability an unsupported or unjustifiable overhead. The results were almost evenly split, with 20 saying yes and 19 saying no.

We then asked attendees why they care about sustainability (beyond just availability / e.g., posting code on GitHub). 50 responded, choosing from the following options:
- 17: Enabling others to go further
- 10: Prevent reimplementation
- 9: Reproducibility
- 9: Increase impact
- 4: Need the software yourself
- 1: Want more users

We asked attendees what they consider the biggest challenge to sustainability, and 52 responded, choosing from the following options:

- 14: Prioritization of other things
- 14: Funding / resources for bug fixes / maintenance
- 9: Time / lack of extra effort
- 7: Finance / money
- 7: Funding / resources for refactoring / rearchitecting
- 1: Ability (inability) to extend / reset codebase

Finally, we asked what strategies attendees are using to increase sustainability for their software. 36 people responded, choosing the following options (multiple responses were allowed):
- 24: Seeking institutional support
- 18: Educating funding agencies that your software is needed
- 10: Seeking letters/testimonials to help attract new grant funding
- 4: Selling support
- 4: Selling higher level functionality / features for money
- 4: Building a paid membership community / consortium

## Conclusions

A few of the key points that came out of the workshop follow. How to address these points is somewhat open, but some of the organizations represented by the panelists are well-suited to start this process. In addition, a newly created umbrella organization, the [Research Software Alliance](), might serve as a means to coordinate these and other organizations to do this.

Both panelists and member of the audience agreed that well-organized code with software engineering practices in place has a better chance of being sustainable and of being widely adopted. Follow-on might include involving the software engineering community in gathering data to support this, and publicizing the results.

While there is general agreement on the need for better software engineering practices and this agreement can be broadened, the path to actually achieving it is not very clear. Certainly, a few elements can seriously improve the situation: education, a robust incentive structure and proper resources, but these need to be supported by hiring institutions, community and professional organizations, and funding agencies.

We may well face a serious loss in digital born information, from the nineties and back, if we can't keep access to the software that created those data. This is a challenge for the repository and curation community.

There appears to be a critical mass of users/contributors that can lead to a software project becoming community supported and a software sustainability infrastructure that connects all experts in the field. More study is needed in this area, and more specific recommendations for good practices to develop such a community and infrastructure are needed. Some organizations that are working in this area include [The Carpentries](), [NumFOCUS](), and [Code for Science & Society]().